\newcommand{\beq}{\begin{equation}}
\newcommand{\eeq}{\end{equation}}
\newcommand{\beqns}{\begin{equation}}
\newcommand{\eeqns}{\end{equation}}
\newcommand{\beqar}{\begin{eqnarray}}
\newcommand{\bs}{\begin{eqnarray*}}
\newcommand{\eeqar}{\end{eqnarray}}
\newcommand{\es}{\end{eqnarray*}}
\newcommand{\beqml}{\begin{mathletters}}
\newcommand{\eeqml}{\end{mathletters}}
\newcommand{\Ptilde}{\tilde{P}}
\newcommand{\SH}{{\cal H}}
\newcommand{\SG}{{\cal G}}
\newcommand{\SP}{{\cal P}}
\documentstyle[aps,preprint]{revtex}
\begin{document}
\draft
\preprint{Ref. SISSA 130/95/CM} 
\title{Diffusion and Trapping on a one-dimensional lattice}
\author{Achille Giacometti} 
\address{International School for Advanced Studies, via Beirut 2-4 
and Sezione INFM di Trieste, I-34013 Trieste, Italy}
\author{K.~P.~N. Murthy }
\address{Theoretical Studies Section, Materials
Science Division, Indira Gandhi Centre for Atomic Research,
Kalpakkam 603 102, Tami Nadu, India}

\date{\today}
\maketitle

\begin{abstract}
The properties of a particle diffusing on a one-dimensional lattice where
at each site a random barrier and a random trap
act simultaneously on the particle are investigated by
numerical and analytical techniques. 
The combined effect of disorder and traps yields a decreasing survival
probability with broad distribution (log-normal).
Exact enumerations, effective-medium approximation  and
spectral analysis are employed. This one-dimensional model shows
rather rich behaviours which were previously believed
to exist only in higher dimensionality.
The possibility of a trapping-dominated
super universal class is suggested.
\end{abstract}
%
\newpage
\narrowtext
\section{Introduction}
\label{sec:intro}
One-dimensional models are widely used in the physics of
disordered systems \cite{LM}. This is because on
one hand they are often good representatives of
higher dimensional models and on the other hand they are
much easier to handle.
In the last decades the importance of investigating simple
diffusion in the presence of disorder and trapping has been
widely appreciated as toy model for complex systems
including migrations of optical excitations \cite{AK},
polymer physics \cite{de Gennes} and diffusion-limited binary
reactions \cite{Mik}. See e.g. \cite{Reviews} and 
\cite{Alexander et al} for an exhaustive review.

From the mathematical viewpoint the analytical solution of the diffusion
problem of particles diffusing before getting completely
absorbed by  permanent traps, has been proven an extremely
difficult task and only an asymptotic solution in presence of
uncorrelated disorder
for the survival probability (to be defined below) could be
given using sophisticated techniques \cite{DV}.

A variation of this problem in the presence of strongly correlated
(percolating) disorder, was also recently numerically 
investigated \cite{G et al,GN} and it
was  observed that
whenever the total survival probability 
is an erratically decreasing function (i.e. detailed balance
is violated) new and unexpected behaviour such
as  enhanced diffusion, breaking of
self-averaging and emergence of Lifshitz tails \cite{G et al,GN} appear.

All these latter phenomena were somehow believed to stem from the
percolating lattice used to model the disorder.

Here we shall investigate a different problem, where
walks are partially and randomly absorbed at {\it each} site
of a one-dimensional lattice
thus leading to a non-conservation of the probability.
From the  physical point of view
this model can mimic the partial absorption of a set of excitons
wandering in mixed crystals.
We shall show that despite the low-dimensionality of the model,
a rich variety of features qualitatively similar to the
ones observed in the trapping in percolating disorder \cite{G et al,GN},
can be found. The low dimensionality of the model however
leaves open the possibility of a full analytical solution
which would extend the results presented here.

Diffusion on a one-dimensional hierarchical lattice \cite{G} and multifractal 
characterization of the escape probability \cite{WGM}
were previously considered in this context. 

Our aim here is then twofold. On one hand we shall 
show that most of the pathological features of the model considered
in Ref. \cite{G et al,GN} are also present in this simplified 
low-dimensional version. On the other hand
the results presented here can be regarded as a complement of
earlier investigations \cite{Alexander et al,SK}
where the total probability is conserved, thus providing a direct test
on the effect of the non-conservation of the probability.

The outline of the paper is as follows. In Sec. \ref{sec:model} we introduce
the model and recall some well known general manipulations of
disordered one-dimensional lattices. Sec. \ref{sec:exact} contains
a numerical solution of the master equation. Sec. \ref{sec:ema}
contains an effective medium approximation to the diffusion, which
is shown to be inadequate. However in Sec. \ref{sec:gp} an heuristic argument
patterned after the Grassberger-Procaccia's similar reasoning for
the Donsker-Varadhan case \cite{DV}, provides an intuitive explanation
of the numerical results. Sec. \ref{sec:spectrum} contains a detailed
numerical investigation of the influence of trapping on the
spectrum and the localization properties.
Finally in Sec. \ref{sec:conclusions} some conclusions are drawn up.
\section{The model}
\label{sec:model}
Consider a particle moving on a one-dimensional lattice
with random barriers and random trapping probability on each site.
The master equation reads:
\beqar \label{ME}
P_{x_0,x}(t+1) &=& \gamma(1-\epsilon_x) P_{x_0,x}(t) +
w_{x,x-1} P_{x_0,x-1}(t)+ w_{x,x+1} P_{x_0,x+1}(t)
\eeqar
for the probability $P_{x_0,x}(t)$ of being at site $x$ at time $t$
having started from site $x_0$ at the initial time $t=0$.
In this notation $w_{x,y}$ is the hopping probability
from site $y$ to site $x$, $\epsilon_x=w_{x-1,x}+w_{x+1,x} < 1$ and
$\gamma \in [0,1]$ is a parameter defining the sojourn probability
which can be continuously tuned from $1$ (no trapping) to $0$ (full trapping).
Thus at each time step a particle at site $x$ can move at
sites $x \pm 1$ with random probability $w_{x,x\pm 1}$, stay at $x$
with sojourn probability $\gamma(1-\epsilon_x)$ and disappear
with probability $(1-\gamma)(1-\epsilon_x)$.
In the following we shall consider only the case of symmetric hopping
probability ($w_{x,y}=w_{y,x}$) and infinite lattice.

As it is well known \cite{LM}, when $\gamma=1$ this model can be mapped into
a variety of other one-dimensional models. 

The (discrete) Laplace transform $\Ptilde_{x_0,x}(\omega)$ of the probability
$P_{x_0,x}(t)$ satisfies an equation of motion which can be cast
in the following Green equation for a Tight Binding (TB) hamiltonian:
\beqar \label{TBGF}
\sum_y (E-\SH(\gamma))_{x,y} G_{x_0,y}(E) &=& \delta_{x,x_0}
\eeqar 
corresponding to the generalized TB hamiltonian
\beqar \label{TBH}
\SH_{x,y}(\gamma) &=& \Omega_x(\gamma) \delta_{y,x} -w_{x,y}
(\delta_{y,x-1}+\delta_{y,x+1})
\eeqar
where we have defined $\Omega_x(\gamma)\equiv \gamma \epsilon_x +1 -\gamma$
and $G_{x_0,x}(E)\equiv-\Ptilde_{x_0,x}(\omega)|_{\omega=-E}$.
Here the (discrete) Laplace transform is defined as:
\beqar \label{DLT}
\Ptilde_{x_0,x}(\omega) &=& \sum_{t=0}^{+\infty}
\frac{P_{x_0,x}(t)}{(1+\omega)^{t+1}}
\eeqar

On the other hand eqn.\ (\ref{ME}) can be written in the
following form:
\beqar \label{TME}
P_{x_0,x}(t+1) &=& \sum_y T_{x,y}(\gamma) P_{x_0,y}(t)
\eeqar
where we have defined the transition matrix
\beqar  \label{TM}
T_{x,y}(\gamma) &=& \left\{ \begin{array}{lll}
 \gamma(1-\epsilon_x) \;\;\; \mbox{if $x=y$} \\
w_{x,y} \;\;\;\; \mbox{if $|x-y|=1$} \\
\;\;\;\;\;\;\; 0 \;\;\; \mbox{otherwise}
              \end{array}
               \right.
\eeqar
It is then easy to check that if $\gamma<1$ (and thus $\sum_x T_{x,y} <1$
for any $x$), then all the eigenvalues $\{\lambda_{\alpha}\}$ of $T$ are
{\it strictly} less then unity, i.e. $|\lambda_{\alpha}|<1$ for any $\alpha$, 
unlike the conserved case ($\gamma=1$) where the fact that
the maximum eigenvalue $\lambda_M$ is non-degenerate and equal
to $1$ irrespective of the (finite) size of the system, is ensured by
the Perron-Frobenius' theorem \cite{P}. Equivalently this means
that the eigenvalues $\{E_{\alpha}\}$ of the Hamiltonian
(\ref{TBH}) are all {\it strictly} positive,
i.e. $E_{\alpha}>0$ for any $\alpha$.

All the results presented in this paper were obtained using the
discrete time equation (\ref{ME}). The continuum time
counterpart of eqn (\ref{ME}) would be, in the Coutinuos Time Random Walk
(CTRW) formulation:
\beqar \label{MEC}
\partial_t P_{x_0,x}(t) &=& {\hat w}_{x,x-1} P_{x_0,x-1}(t)+
{\hat w}_{x,x+1} P_{x_0,x+1}(t)-
{\hat \epsilon}_x P_{x_0,x}(t) -{\hat \beta}_x P_{x_0,x}(t)
\eeqar
where we have indicated with ${\hat \beta}_x$ the absorption rate at site $x$
and used the tildas to indicate the quantities which are rates.
By balancing the gain, loss and sojourn terms to unity, it is easy
to see that ${\hat \beta}_x=(1-\gamma)(1-{\hat \epsilon}_x)$.
However it should be stressed that eq. (\ref{MEC}) is {\it not}
the continuum limit of eq. (\ref{ME}). 

\section{Exact enumeration}
\label{sec:exact}
We have considered a distribution of of the disorder given by:
\beqar \label{disorder}
\rho(w) &=& 2^{1-\alpha}(1-\alpha) w^{-\alpha} \theta(w) \theta(1/2-w)
\eeqar
where $\alpha \in (-\infty,1)$. By varying the parameter $\alpha$ we can
pass from {\it weak} disorder ($\alpha <0$) to {\it strong} disorder
($\alpha \to 1$). The case $\alpha=0$ corresponds to a uniform distribution
and it is {\it marginal} in the sense that the inverse 
first moment is logarithmically
divergent. We shall indicate with a overbar the average over the disorder
(\ref{disorder})  which is assumed to be {\it quenched}.
The simplest quantity that one would like to compute is the mean-square
displacements:
\beqar \label{MSD}
<x^2(t)> &=& \overline{
\frac{\sum_x (x-x_0)^2 P_{x_0,x}(t)}{\sum_x P_{x_0,x}(t)}}
\eeqar
where it should be noted that the denominator must be included
since the total probability is {\it not} conserved. The disorder
average of the latter is called {\it survival probability} $P_s(t)$
and it is a decreasing function of time having chosen the initial
conditions $P_s(0)=1$. The return probability 
\beqar \label{return}
P_0(t) &=& \overline{\frac{P_{x_0,x_0}(t)}{\sum_x P_{x_0,x}(t)}}
\eeqar
is another interesting quantity to look and it is again different
from the {\it survival return} probability 
$P_s^0(t) =\overline{P_{x_0,x_0} (t)}$  in view of the non-conservation
of the probability.
A suitable normalization procedure \cite{G et al} is to be used
in order to avoid a quick underflow of the survival probability.

We numerically solved the master equation exactly up to $t=1000$
for three representative  values of the strength of the disorder,
namely $\alpha=-0.5$ (Weak disorder), $\alpha=0$ (Marginal disorder)
and $\alpha=0.5$ (Strong disorder). The lattice size was chosen
sufficiently large (up to $N=2^{18}$) to avoid finite size effect
due to the boundaries. Our estimates are based on an average
of $3$ samples of $1000$ different configurations each.
Errors are statistical. 

The case $\gamma=1$ was previously studied in Ref. \cite{Alexander et al}
and our numerical results are in perfect agreement with the
one reported there. Indeed we find:
\beqar \label{nu}
<x^2(t)> &=& t^{2 \nu}
\eeqar
where $\nu$ is the correlation length exponent, and
\beqar \label{dtilde}
P_0(t) &=& t^{-d_s/2}
\eeqar
where $d_s$ is the return probability exponent. Our estimates are
$2 \nu =0.99 \pm 0.01$ and $0.68 \pm 0.01$ for $\alpha=0,0.5$
respectively which compares well with the expected values 
$1$ and $2/3$.

The case $\gamma=0$, corresponding to
a full trapping, appears to be completely {\it trapping-dominated}. 
Indeed here the sample-to-sample fluctuations  are enormous and they completely
rule the diffusion. In order to have a feeling for this, we
plotted in Fig. \ref{fig1} the mean-square displacement 
for a single configuration obtained with the same initializing
seed, in the case $\gamma=1$ and
$\gamma=0$ for various strengths of the disorder. 
The staircase behaviour which can be observed in
the case $\gamma=0$ is a consequence of the non-conservation
of the probability. A similar effect occurs in the return probability.
This rather peculiar feature was already observed in the model of 
Ref. \cite{G et al}. Here however it is noteworthy the extremely weak
dependence of the behaviour case from the strength of the disorder.

Upon disorder average we find, in the $\gamma=0$ case, that the first
moment ($<x(t)>$) is zero as expected and the second moment $<x^2(t)>$
and the return probability $P_0(t)$ follow the behaviour
(\ref{nu}) and (\ref{dtilde}) respectively.

Our best estimate for the exponents $\nu$ and $d_s$ are
reported in Table \ref{table1}. As mentioned above, the very similar 
result for the
three case of $\alpha$ is somewhat surprising, but it is in agreement
with the aforementioned interpretation.
Although the three values are not all
within the numerical errors, we cannot rule out the possibility that
the universality class be {\it independent} on the disorder choice.
We also checked that all the intermediate cases $0<\gamma<1$
behave as in the $\gamma=0$ case after a transient time
which depends on the value of $\gamma$.
It should be stressed that $d_s$ bears the meaning of {\it spectral
dimension} \cite{Alexander et al} only in the case $\gamma=1$.

Although we monitored the behaviour of the survival probability
and found that it decays as a stretched exponential (i.e. with an 
argument for the exponential which is a power of the time
less than one), we also
found that the numerical value of the exponent is not very
easy to pin down. This was to be expected: a similar feature
occurs in the Donsker-Varadhan problem \cite{havlin et al}.

In view of these results, one expects both the
survival probability and the survival return probability
to be {\it non-self-averaging} quantities. We investigated
the full probability distributions of both quantities.
We find (see Fig. \ref{fig2}) that they rather accurately 
follow a log-normal distribution:
\begin{eqnarray} \label{log-normal}
\SP[X(t)]&=&\frac{1}{X\sqrt{2 \pi \sigma_t^2}} \exp[\frac{-(\ln X
-\lambda_t)^2} {2 \sigma_t^2}]
\end{eqnarray}
where $X(t)=P_{x_0,x_0}(t),\sum_x P_{x_0,x}(t)$. Here $\lambda_t$
and $\sigma_t^2$ are the mean and the variance of the distribution
respectively. If asymptotically ($t>>1$) it happens that
$\sigma_t^2>> \lambda_t$ then self-averaging is {\it broken}
(see discussion in Ref.\cite{G et al}).
Both these quantities can be computed directly as first ($\lambda_t$) and
second ($\sigma_t^2$) moments of the distribution but also indirectly
by fitting the evolution of the log-normal function at various values of
$t$. We find the following behaviour for $t>>1$:
\beqar \label{chi}
\sigma_t^2 & \sim t^{2 \chi}
\eeqar
for the survival probability (and this defines the "free-energy" 
exponent $\chi$) and a similar behaviour for the survival return
probability (which defines the analog exponent $\chi_0$).
Our best estimates for these exponents is reported in Table \ref{table2}.
On the other hand we find that asymptotically $\lambda_t \sim t$ both
for the survival and survival return probability. Since
in all cases $2 \chi >1$ the self-averaging property is broken 
\cite{G et al}.

As a final remark we computed higher moments ($<x^4(t)>$, $<x^6(t)>$, etc)
in the attempt to find signatures of multifractality
as suggested by the results of Ref. \cite{WGM}. We found that all the moments
were related to the second one {\it irrespectively} of the
strength of the disorder, thus ruling out the possibility
of multifractal behaviour in the size of the walk.

\section{Effective Medium Approximation}
\label{sec:ema}
In this section we will tackle the problem of solving eq (\ref{ME})
by using an Effective Medium Approximation (EMA). Although this
approximation is known to fail in many
situations \cite{Reviews}, it nevertheless gives extremely accurate
results for the problem described by (\ref{ME}) with $\gamma=1$
\cite{Alexander et al}. Here we shall carry out the
successful recipe given in \cite{Alexander et al} to the other extreme,
namely the case of full trapping ($\gamma=0$). 

The equation of motion for the Laplace transform $\Ptilde_{x_0,x}(\omega)$
can be written in the form (taking $x_0=0$ for concreteness)
\beqar \label{Laplace}
\SG_{0,x}(\omega) &=& \mu_{x-1}(\omega) \SG_{0,x-1}(\omega)+
\mu_{x} (\omega)\SG_{x+1}(\omega) +\delta_{x,0}
\eeqar
where $\SG_{0,x}(\omega)=(1+\omega) \Ptilde_{0,x}(\omega)$ and where
we have defined $\mu_{x} (\mu_{x-1})=w_{x,x+1}(w_{x-1,x})/(1+\omega)$.
For $x>0$ and $x<0$ one can introduce the following new fields
\beqar \label{phi}
\phi_x^{+}(\omega) = 
\frac{\mu_{x-1}(\omega) \SG_{0,x}(\omega)}{\SG_{0,x-1}(\omega)} 
&\hspace{0.5in}& \phi_x^{-}(\omega) = 
\frac{\mu_{x}(\omega) \SG_{0,x}(\omega)}{\SG_{0,x-1}(\omega)} 
\eeqar
respectively. In this way the problem is reduced to first-order
and can be solved by continued fractions. For $x>0$ one finds
the following recursions:
\beqar \label{phi+}
\phi_x^{+}(\omega) &=& 
\frac{\mu_{x-1}^2(\omega)}{1-\phi_{x+1}^{+}(\omega)} 
\eeqar
A similar equation can be found for $x<0$. It should be noted that
the support for the $\phi^{+,-}$ have to be constrained in
such a way that eq (\ref{phi+}) be sensible.
From eq. (\ref{Laplace}) for the case $x=0$ and using eq. (\ref{phi+})
and the corresponding for $x<0$ one easily finds that
\beqar \label{G_0}
\SG_{0,0}(\omega) &=& \frac{1}{1-\phi_{+1}^{+}(\omega)-\phi_{-1}^{-}(\omega)}
\eeqar
The disorder average of eq. (\ref{G_0}) thus gives:
\beqar \label{G_0ave}
\overline{\SG_{0,0}(\omega)} &=& \int \; d \phi_{+} \int \; d \phi_{-}
\Pi_{\omega}^{+}(\phi_{+}) \; \Pi_{\omega}^{-}(\phi_{-}) 
\frac{1}{1-\phi_{+}-\phi_{-}} \; \theta(1-\phi_{+}-\phi_{-})
\eeqar
where the distributions $\Pi_{\omega}^{+}(\phi_{+})$ and 
$\Pi_{\omega}^{-}(\phi_{-})$ are given by
\beqar \label{Pi}
\Pi_{\omega}^{+}(\phi_{+}) &=& \int \; d \mu \; \rho (\mu) \; 
\int \; d \phi_{+}^{\prime} \; \Pi_{\omega}^{+}(\phi_{+}^{\prime})
\; \delta(\phi_{+} - \frac{\mu^2}{1-\phi_{+}^{\prime}})
\eeqar
and similarly for $\Pi_{\omega}^{+}(\phi_{-})$.
Now comes the EMA approximation:
\beqar \label{EMA}
\Pi_{\omega}^{+}(\phi_{+})=\Pi_{\omega}^{-}(\phi_{-})= 
\delta(\phi-\phi_e(\omega))
\eeqar
where $\phi_e(\omega)$ is an {\it effective} field to be found
self-consistently. It is immediate to show that the self-consistency
equation decouples and its solution is given by:
\beqar \label{EF} 
\phi_e(\omega) &=& \frac{1-\sqrt{1-4 \mu_a^2(\omega)}}{2}
\eeqar
($\mu_a(\omega)=\mu_a/(1+\omega)$) provided that the second moment 
of the distribution 
\beqar \label{second}
\mu_a^2 &\equiv& 
\frac{\int \; d \mu \; \mu^2 \; \rho(\mu)}{\int \; d \mu \; \rho(\mu)}
\eeqar
exists. As a consequence:
\beqar \label{G_02}
\overline{\SG_{0,0}(\omega)} &=& \frac{1}{\sqrt{1-4 \mu_a^2(\omega)}}
\eeqar
Due to conditions imposed on the transition rates, we have that
$\mu_a \leq 1/2$. If equality holds then the result is identical to
the one obtained in the absence of disorder. If however $\mu_a <1/2$
then by antitransforming back to direct domain (in the continuum 
approximation), it is easy to show 
that the {\it survival return } probability behaves as:
\beqar \label{survival return}
P_s^0(t) &=& e^{-t} I_0(2 \mu_a t)
\eeqar
where $I_0(z)$ is the 0-th order Bessel function.  Upon asymptotic
expansion \cite{AS} the leading behaviour for $t>>1$ is
\beqar \label{asymptotic}
P_s^0(t) &=& \frac{e^{-(1-2\mu_a)t}}{\sqrt{4 \pi \mu_a t} }
[1+\frac{1}{16 \mu_a t} + 0(\frac{1}{t^2})]
\eeqar
The result is {\it qualitatively} different from the numerical
indication. This was to be expected in such \cite{Reviews} EMA may not be 
capable of capturing the nuances of systems where large sample-to-sample
fluctuations are present.

\section{An Heuristic Argument}
\label{sec:gp}
In this section we shall present an intuitive argument yielding
a stretched exponential for the survival probability in {\it qualitative} 
agreement with our numerical findings \cite{Kehr}.
The argument is based on a similar one given by Grassberger and Procaccia
\cite{DV} for the Donsker-Varadhan problem, and provides a plausible
explanation of the irrelevance of the choice of the distribution
for the hopping probability $\rho(w)$.

The main idea is that the maximum contribution to the
survival probability is coming from rare regions where
the hopping probabilities are all very close to $1/2$, say
in the interval $[1/2-\epsilon,1/2]$ (with $0<\epsilon <<1/2$)
for the sake of the argument.
If $s$ is the number of sites of one of these regions, the 
typical number of steps $\tau$ necessary to explore the
region is $\tau \sim s^{1/\nu}$. If there were no leaking of
probability the decay of the survival probability in this
region $P(t,s)$ would be exponential. However, during the time $\tau$,
there is a loss of
probability of order $\epsilon^{\tau}$. The probability ${\cal P}(s)$
of finding such a region is $(2\epsilon)^s$. Therefore
the survival probability is expected to have the form:
\beqar \label{survival}
P_s(t) &=& \int_{0}^{+\infty} \;\; ds \; {\cal P}(s) P(t,s) \sim
\int_{0}^{+\infty} \;\; ds \; \exp[-f_t(s)]
\eeqar
where for $\epsilon <<1$ we have that
\beqar \label{argument}
f_t(s) &=& \frac{t}{s^{1/\nu}}+\lambda s^{1/\nu}+\lambda s
\eeqar
where $\lambda=|\ln \epsilon|$. The integral can be carried out using the
steepest descent method which yields the following equation
\beqar \label{saddle point}
t &=& \lambda s_0^{2/\nu} +\lambda \nu s_0 ^{\frac{1+\nu}{\nu}}
\eeqar
for the saddle point $s_0$. Since $\nu<1$, at the leading order
for $t>>1$, the solution is:
\beqar \label{solution}
s_0 &\sim& (1/\lambda)^{\nu/2} t^{\nu/2} 
\eeqar 
Therefore, upon substitution in eq. (\ref{survival}) we find at the leading
order in $t>>1$:
\beqar \label{final}
P_s(t) &\sim& \exp(-C \sqrt{t})
\eeqar
where $C$ is a constant, which is different from the Donsker-Varadhan 
case where the argument of the exponential is $\sim -t^{1/3}$.

This argument seems to suggest the independence of the strength of
the disorder which is irrelevant in the regions in the neighbourhood
of $1/2$. Clearly the argument can be extended to higher dimensions
and self-similar lattices.

\section{Spectral Analysis}
\label{sec:spectrum}
\subsection{Density of States}
\label{subsec:dos}
We shall here study the spectral properties of the TB equation
associated to (\ref{TBGF}):
\beqar \label{TBE}
\Omega_x(\gamma)  \psi_x(E) -w_{x,x-1} \psi_{x-1}(E)
-w_{x,x+1} \psi_{x+1}(E) &=& E \psi_x(E)
\eeqar
by studying the density of states and the localization properties.
For $\gamma=1$ the result is well known \cite{Hori}. The Hamiltonian
is positive definite as one can immediately check by defining the
following creation 
\beqar \label{creation}
(a^{\dagger} \psi)_x &=& \sqrt{w_{x,x-1}} \psi_{x-1} -\sqrt{w_{x,x+1}} \psi_x
\eeqar
and destruction 
\beqar \label{destruction}
(a \psi)_x &=& \sqrt{w_{x+1,x}} \psi_{x+1} -\sqrt{w_{x,x+1}} \psi_x
\eeqar
operators and noting that $\SH(1)=a^{\dagger} a$ \cite{supersymmetry}.
Then for any $\gamma<1$ we have that, using $ <\psi|\varphi>$ to indicate
the usual scalar product:
\beqar \label{energy}
<\psi|\SH(\gamma)|\psi> \;\; > \;\; <\psi|\SH(1)|\psi> \;\; \ge 0
\eeqar
and thus the Hamiltonian $\SH(\gamma)$ is strictly positive
for any $\gamma<1$.

A well known and efficient way  to compute the spectrum of
a one-dimensional disordered model is by means of the 
{\it node-counting} theorem \cite{Hori}.
By defining
\beqar \label{U}
U_x(E) &=& w_{x-1,x} \frac{\psi_x(E)}{\psi_{x-1}(E)}
\eeqar
the following recursion can be easily found from (\ref{TBE}):
\beqar \label{recursions U}
U_{x+1}(E) &=& (\Omega_x(\gamma)-E) - \frac{w_{x,x-1}^2}{U_x(E)}
\eeqar 
By calculation the number of times the quantity $U_{x}(E)$ changes
sign, the integrated density of states
\beqar \label{IDOS}
M(E) &=& \int_{-\infty}^{E} \; d E' \; N(E')
\eeqar
(rather that the density of states $N(E)$) can be easily computed.
The validity of the numerical procedure can be tested against
the periodic case where the exact result 
\beqar \label{exact}
M(E) &=& \frac{1}{2}-\frac{1}{\pi} \arcsin(1-E)
\eeqar
with $E \in [0,2]$, can be easily derived.
The case $\gamma=1$ was analytically solved by Stephen and Kariotis
\cite{SK}. They found in the $E \to 0$ limit:
\beqar \label{SK1}
M(E) &=&  \left\{ \begin{array}{ll}
	   E^{\frac{1-\alpha}{2-\alpha}} \;\; \mbox{for $0<\alpha<1$} \\
	   E^{1/2}/\sqrt{|\ln \sqrt{E}|} \;\; \mbox{ for $\alpha=0$}
	   \end{array}
                \right.
\eeqar
Our numerical results for $\alpha=0.5$ and $\alpha=0$ reproduce very
well the theoretical prediction. Indeed we find $0.34 \pm  0.01$
and $0.45 \pm 0.01$ for the two cases where the theoretical
values are $1/3$ and $1/2$ respectively. The $\alpha=0$ case
is less precise due to the logarithmic corrections.
Our results are based on an average of $5$ different configurations
with $N=2^{18}$ number of sites. We found that the the results were
changing very little from one configuration to another (this is
tantamount to say that $N(E)$ is self-averaging) and thus
this average was sufficient for our purposes. 

The behaviour for $\gamma=0$ is {\it qualitatively } different
(see Fig.\ref{fig3}).
We find for $E \to 0$ (see Fig. \ref{fig4}) a Lifshits tail singularity:
\beqar \label{lifshits}
M(E) &=& \exp(-1/E)
\eeqar
for all three cases $\alpha=-0.5,0,0.5$. As expected on the basis
of the result of Sec. \ref{sec:exact}, all three values
of the disorder give the same quantitative results. Fluctuations in
the tail region grow as the disorder becomes stronger.
We have been unable to give a theoretical derivation for this
behaviour. It should be noted that the EMA of Sec. \ref{sec:ema}
yields:
\beqar \label{DOS_EMA}
N(E) &=& \frac{1}{\pi} \frac{1}{\sqrt{2 E-E^2-c}}
\eeqar
for $E \in [1-2\mu_a,1+2\mu_a]$ and $0$ otherwise with $c=1-4 \mu_a^2$.
With $c=0$ this is just the well known result for a periodic one-dimensional
lattice. Again the EMA fails to give qualitatively correct predictions.
\subsection{Lyapunov exponent}
\label{subsec:ljap}
Eq.(\ref{TBE}) can be cast in a transfer matrix form
\beqar \label{transfer}
\Psi_{x+1}(E) &=& T_x(E) \Psi_x(E)
\eeqar
where we have defined the matrices:
\beqar \label{transmission}
 T_x(E) &=& 
\: \left( \begin{array}{cc}
0 & 1 \\
-\frac{w_{x,x-1}}{w_{x,x+1}} & \frac{\Omega_x(\gamma)-E}{w_{x,x+1}}
\end{array} \;\; \right)
\eeqar
and
\beqar \label{field}
\Psi_x(E) &=& 
\: \left( \begin{array}{c}
\psi_{x-1}(E) \\
\psi_{x}(E) 
\end{array} \;\; \right)
\eeqar
Upon iteration eq. (\ref{transfer}) becomes, if $N$ is the number
of sites:
\beqar \label{iteration}
\Psi_N(E) &=& (\prod_{x=0}^{N-1} T_x(E)) \Psi_0(E)
\eeqar
where we have taken the initial vector as:
\beqar \label{initial}
\Psi_0(E) &=&
\: \left( \begin{array}{c}
0 \\
1
\end{array} \;\; \right)
\eeqar
The Lyapunov exponent is then defined as:
\beqar \label{lyapunov}
\gamma(E) &=& \lim_{N \to \infty} \frac{1}{N} \ln ||\Psi_N(E) ||
\eeqar
Again we tested our algorithm in the case $\gamma=1$ where an
exact result \cite{SK} gives for $E \to 0$:
\beqar \label{SK2}
\gamma(E) &=&  \left\{ \begin{array}{ll}
	   E^{\frac{1-\alpha}{2-\alpha}} \;\; \mbox{for $0<\alpha<1$} \\
	   E^{1/2}/\sqrt{|\ln \sqrt{E}|} \;\; \mbox{ for $\alpha=0$}
	   \end{array}
                \right.
\eeqar
Our results are $0.34 \pm 0.01$ (to be compared with $1/3$) and
$0.53 \pm 0.02$ (to be compared with $1/2$).
For $E=0$ the Lyapunov exponent is zero as it should since it
corresponds to the stationary state of the master equation (\ref{ME})
\cite{kampen}.

In the $\gamma=0$ case instead we found that $\gamma(0)=c(\alpha)$
that is the ground state is {\it localized} (see Fig. \ref{fig5}). 
The spectrum is symmetric around $E=1$  and this stems
from the fact that the matrix $\SH(0)-1$ is traceless.
Again all the intermediate cases ($0<\gamma<1$) follow this behaviour
with a shift of the center of the spectrum which depends on the
value of $\alpha$.

\subsection{Exact diagonalization}
\label{subsec:exact2}
As a cross checking for the results of Sec.\ref{subsec:dos} and 
\ref{subsec:ljap},
we direct diagonalized the matrix in (\ref{TBH}) both for the
case $\gamma=1$ and $\gamma=0$ by using standard IMSL package routines
for small matrices ($N=100$).
Although a direct quantitative comparison with the previous results
is out of discussion since much higher sizes would be necessary,
these results provide a qualitative understanding of the 
main difference in the spectrum of (\ref{TBH}) in the
cases $\gamma=1$ and $\gamma=0$.
In Fig. \ref{fig6} the spatial dependence of the eigenvector
$\psi_x(E)$ is displayed for $\gamma=0,1$ for an energy in the
lower (a), central (b) and upper (c) part of the spectrum. The
chosen value for the strength of the disorder was $\alpha=0$, but
no qualitative change in the behaviour is found for other
values of the disorder in the relevant case $\gamma=0$.
Correctly the ground state $E=0$ for $\gamma=1$ is extended
as we argued before.

The $\gamma=0$ ground state is very localized consistently with
our previous findings from the Lyapunov exponent.

Another general probe to test the localization properties is
to compute the Inverse Partecipation Ratio (IPR) 
\beqar \label{IPR}
p(E) &=& \frac{1}{N} \frac{(\sum_x |\psi_x(E)|^2)^2}{\sum_x|\psi_x(E)|^4}
\eeqar
where $E$ and $\psi_x(E)$ are the eigenvalues and eigenvectors of
eq.  (\ref{TBE}) and $N$ is the number of sites on the lattice. 
In this notation, then a state $E$ is extended
when $p(E) \sim 1$ while is localized whenever $p(E) \sim 1/N <<1$.
The results are shown in Fig. \ref{fig7} and are consistent with the
previous picture. 

\section{Conclusions}
\label{sec:conclusions}
In this paper we have presented a detailed investigation of the
properties of a one-dimensional disorder model for diffusion where
at each site a fraction of the initial particles could disappear
with random probability depending on the random adjacent barriers.
Our work combined numerical and analytical techniques. 
The main results of this investigations can be summarized as follows.
\begin{description}
\item{1)} We showed that the trapped case ($\gamma<1$) is
qualitatively different from the conserved ($\gamma=1$) counterpart and
that a simple mean-field type of approach is not able to
capture the large fluctuation introduced by the non-conservation
of the probability.
\item{2)} We found that the trapped case appears to be completely
disorder-dominated, thus suggesting a super-universal behaviour
independent on the disorder (which is {\it not} the case
for the conserved problem). 
We provided an intuitive explanation based on an heuristic derivation
of the survival probability of why this may occur.

We believe that more analytical work would be necessary for a 
complete understanding of the particular features
appearing in the model considered in this paper. Work in this
direction is ongoing and will be reported in the future. 
\end{description}

\acknowledgments
Enlightening discussion with Klaus Kehr and Hisao Nakanishi are
gratefully acknowledged. We also thank Klaus Kehr and the referee 
for having suggested the heuristic argument of Sec. \ref{sec:gp}.
The work of AG was supported by the 
{\it Human Capital and Mobility} program under the contract ERB4001GT932058.

\begin{figure}
\caption{Log-Log plot of the mean square displacement $<x^2(t)>$
for a single configuration in the cases $\gamma=1$ (no traps)
and $\gamma=0$ (full trap) for representative values of
the disorder ($\alpha=-0.5,0,0.5$). The $\gamma=0$ case appears to
be independent of the strength of the disorder parameter $\alpha$.}
\label{fig1}
\end{figure}
\begin{figure}
\caption{Distribution density probability for  the survival probability
for time steps $t=200,400,600,800,1000$. The full lines correspond
to the fitted log-normal distribution of the form 
(\protect{\ref{log-normal}}). Shown are the cases $\alpha=-0.5,0,0.5$
corresponding to (a),(b) and (c) respectively.}
\label{fig2}
\end{figure}
\begin{figure}
\caption{Integrated density of states on the whole spectrum
$E \in [0,2]$ for $\gamma=1$ (a) and $\gamma=0$ (b) for the
same strengths of disorder as above.}
\label{fig3}
\end{figure}
\begin{figure}
\caption{Plot of the logarithm of the integrated density of states 
$\ln M(E)$ versus $1/E$ for $E \to 0$ in the case $\gamma=0$ (full trapping)
for $\alpha=-0.5,0,0.5$. The behaviour is qualitatively different
from the case $\gamma=1$ where $M(E)$ is a power law in $E$.}
\label{fig4}
\end{figure}
\begin{figure}
\caption{Same thing as in Fig. \protect{\ref{fig3}}, for the Lyapunov exponent
$\gamma(E)$. A non-zero value
of $\gamma(E)$ indicates that the eigenvalue $E$ is localized.}
\label{fig5}
\end{figure}
\begin{figure}
\caption{Spatial dependence of the eigenvectors $\psi_x(E)$ corresponding
to the lower $(E_M)$ (a), central $(E_0)$ (b) and upper $(E_m)$ 
(c) part of the spectrum for the case $\alpha=0$ and $\gamma=0,1$.}
\label{fig6}
\end{figure}
\begin{figure}
\caption{Inverse Partecipation Ratio (IPR) defined in 
(\protect{\ref{IPR}}) as function of the eigenvalue $E$ for $\alpha=0$
and $\gamma=0,1$. Values close to $1$ indicate that the state is
extended, while values close to $0$ ($\sim 1/N$) mean that
the corresponding eigenvalue $E$ is localized.}
\label{fig7}
\end{figure}
\begin{table}
\caption{Summary of the exponents $\nu$ and $d_s$ for the $\gamma=0$
case in three representative values of the strength of the 
disorder. }
\begin{tabular}{lcc}
\multicolumn{1}{l}{$\alpha$}&
\multicolumn{1}{c}{$2 \nu $}&
\multicolumn{1}{c}{$d_s/2$}\\
\hline
$-0.5$&$1.24 \pm 0.01$&$0.57 \pm 0.01$ \\
$0$&$1.25 \pm 0.01$&$0.59 \pm 0.01$ \\
$0.5$&$1.26 \pm 0.01$&$0.60 \pm 0.01$ \\
\end{tabular}
\label{table1}
\end{table}
\begin{table}
\caption{Summary of the exponents $\chi$ and $\chi_0$ 
defined in the text.
The label (D) and (I) mean direct evaluation and from the
log-normal distribution, respectively. The values corresponding
to a dash were considered unreliable due to the presence of
a strong curvature in the preasymptotic regimes.}
\begin{tabular}{lcccc}
\multicolumn{1}{l}{$\alpha$}&
\multicolumn{1}{c}{$ \chi$(D)}&
\multicolumn{1}{c}{$ \chi_0$(D)}&
\multicolumn{1}{c}{$ \chi$(I)}&
\multicolumn{1}{c}{$ \chi_0$(I)} \\
\hline
$-0.5$&$0.70 \pm 0.01$&$0.72 \pm 0.01$
&$ 0.72 \pm 0.01 $&$0.70 \pm 0.01$ \\
$0$&$0.73 \pm 0.01$&$-$
&$ 0.72 \pm 0.01 $&$0.70 \pm 0.01$ \\
$0.5$&$-$&$-$
&$ 0.72 \pm 0.01 $&$0.70 \pm 0.01$ \\
\end{tabular}
\label{table2}
\end{table}
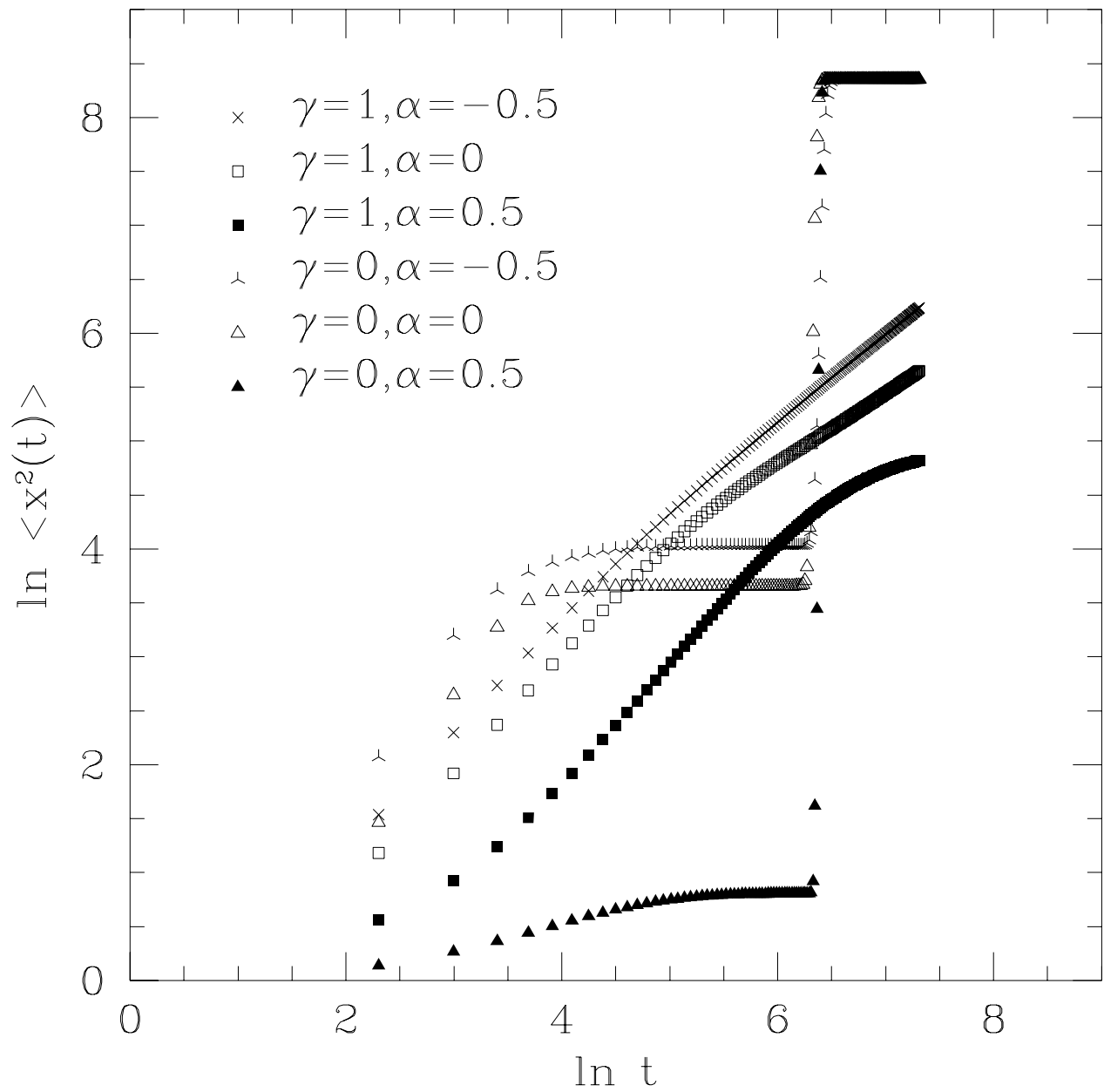
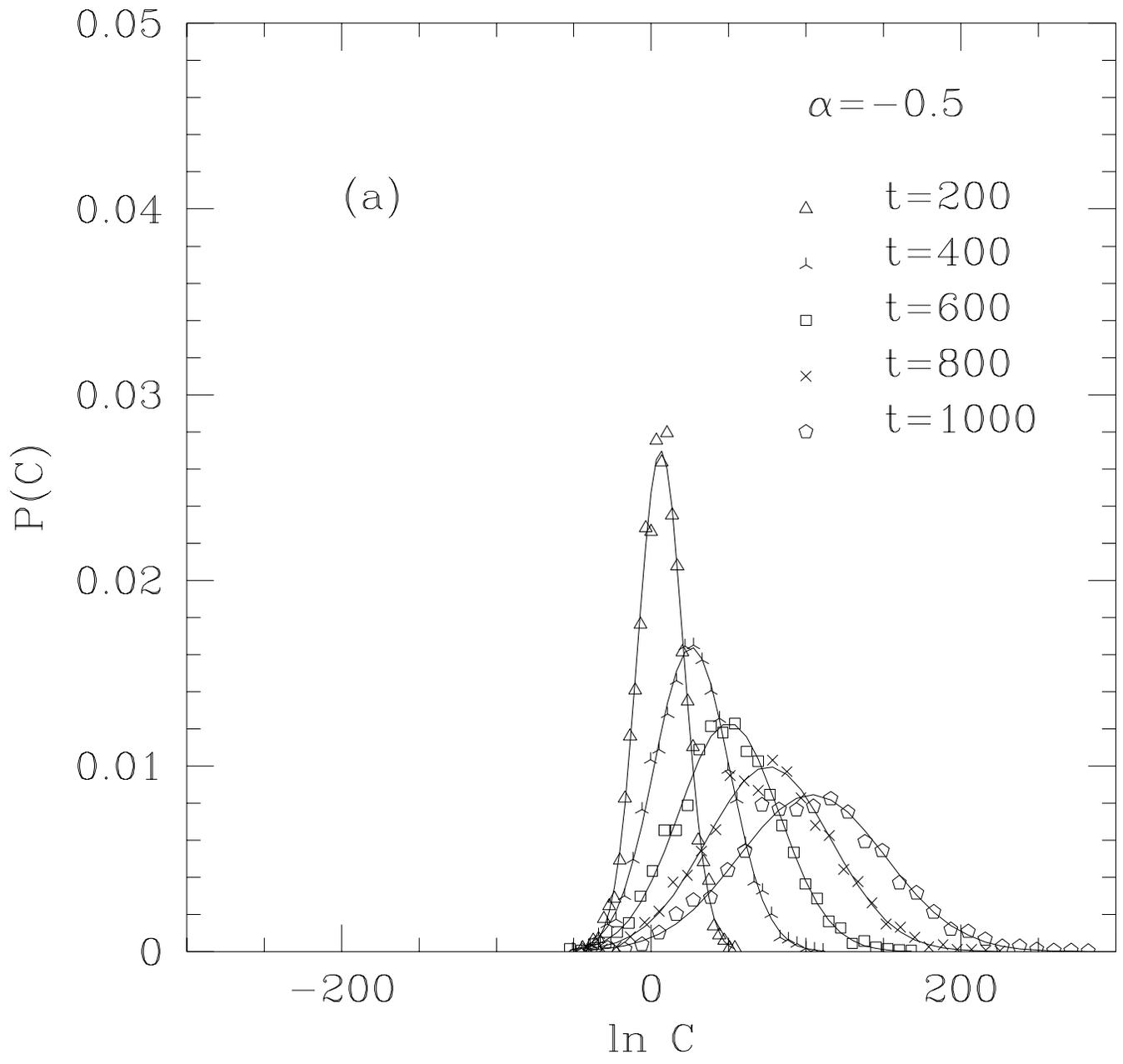
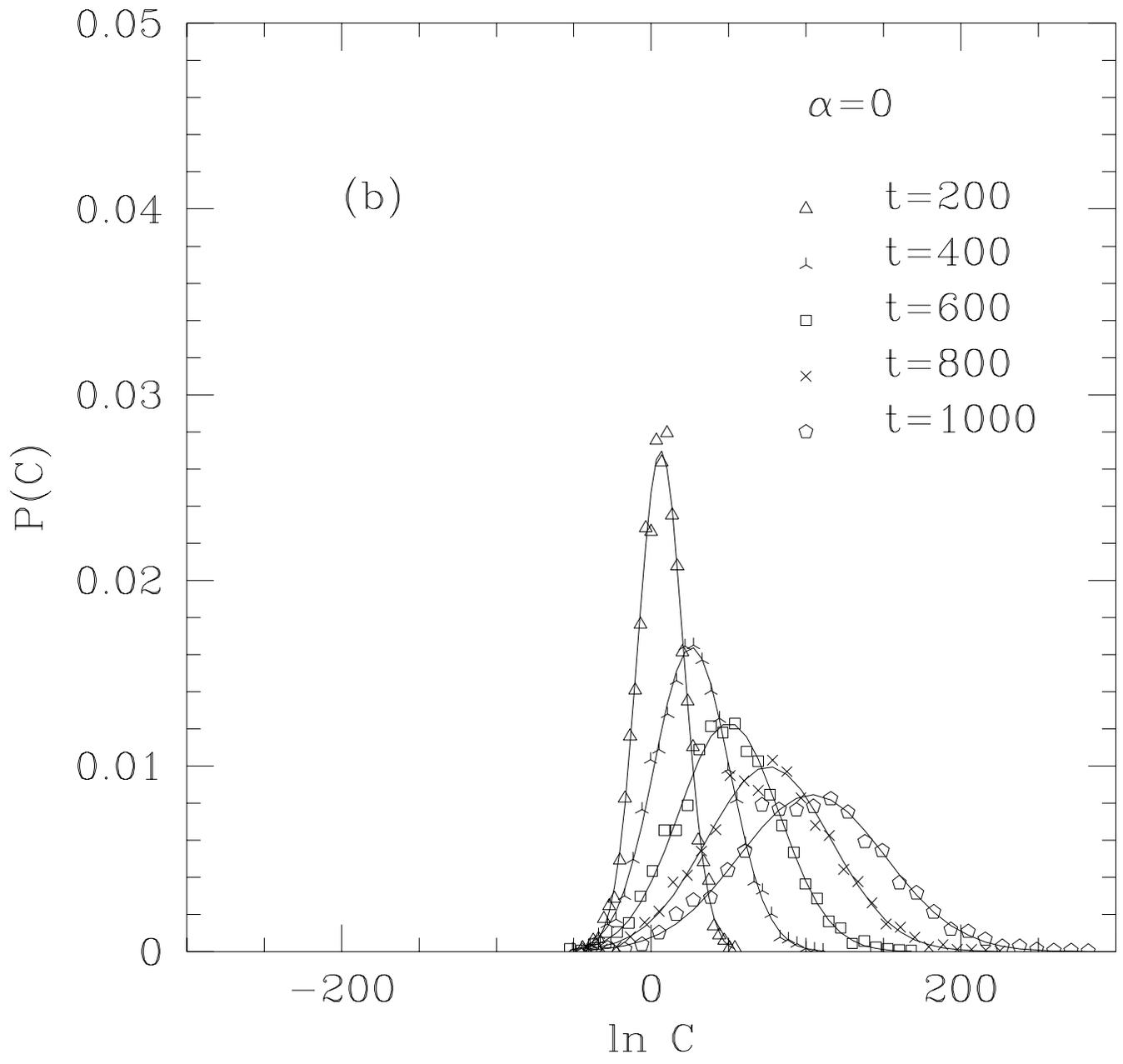
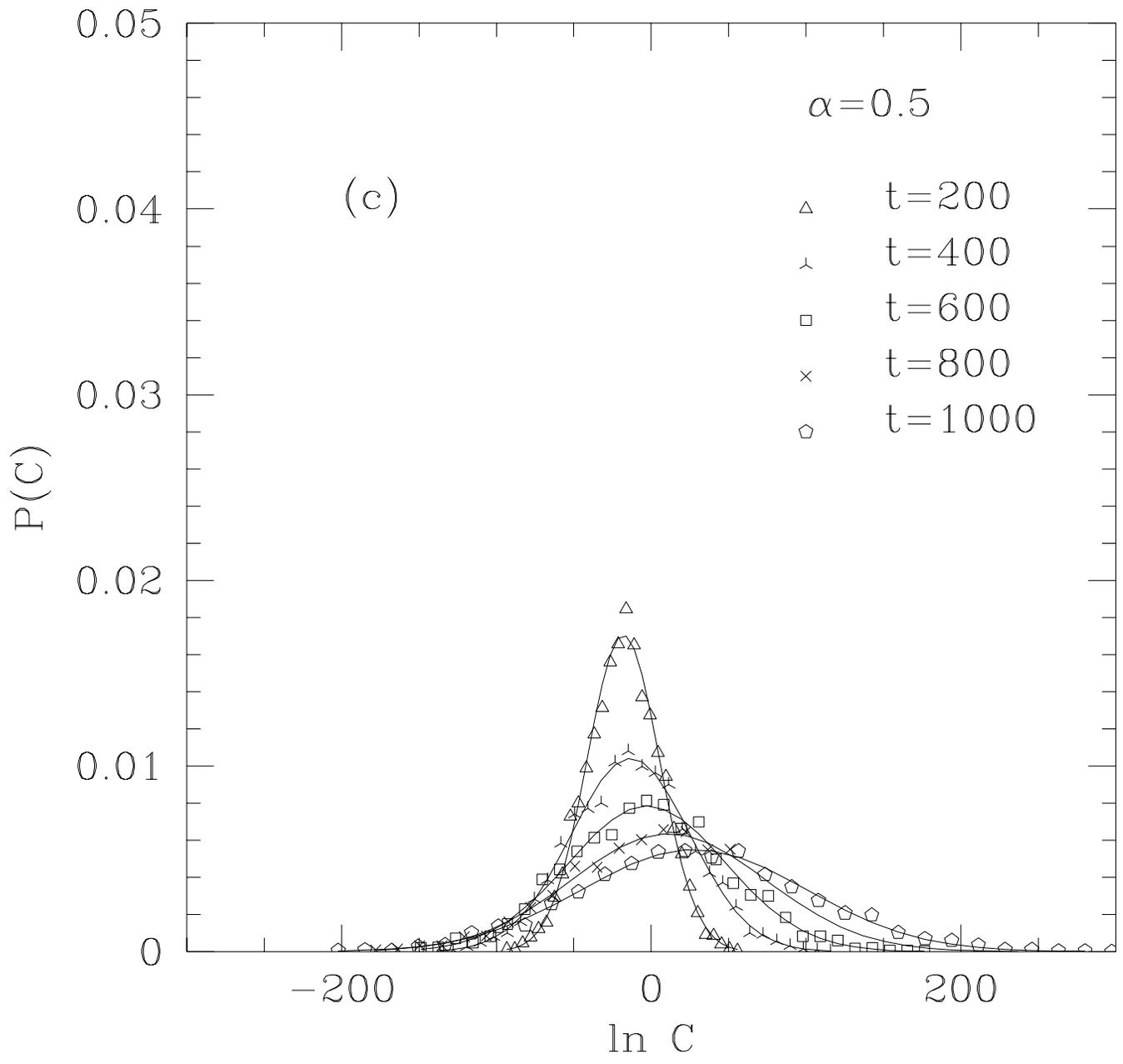
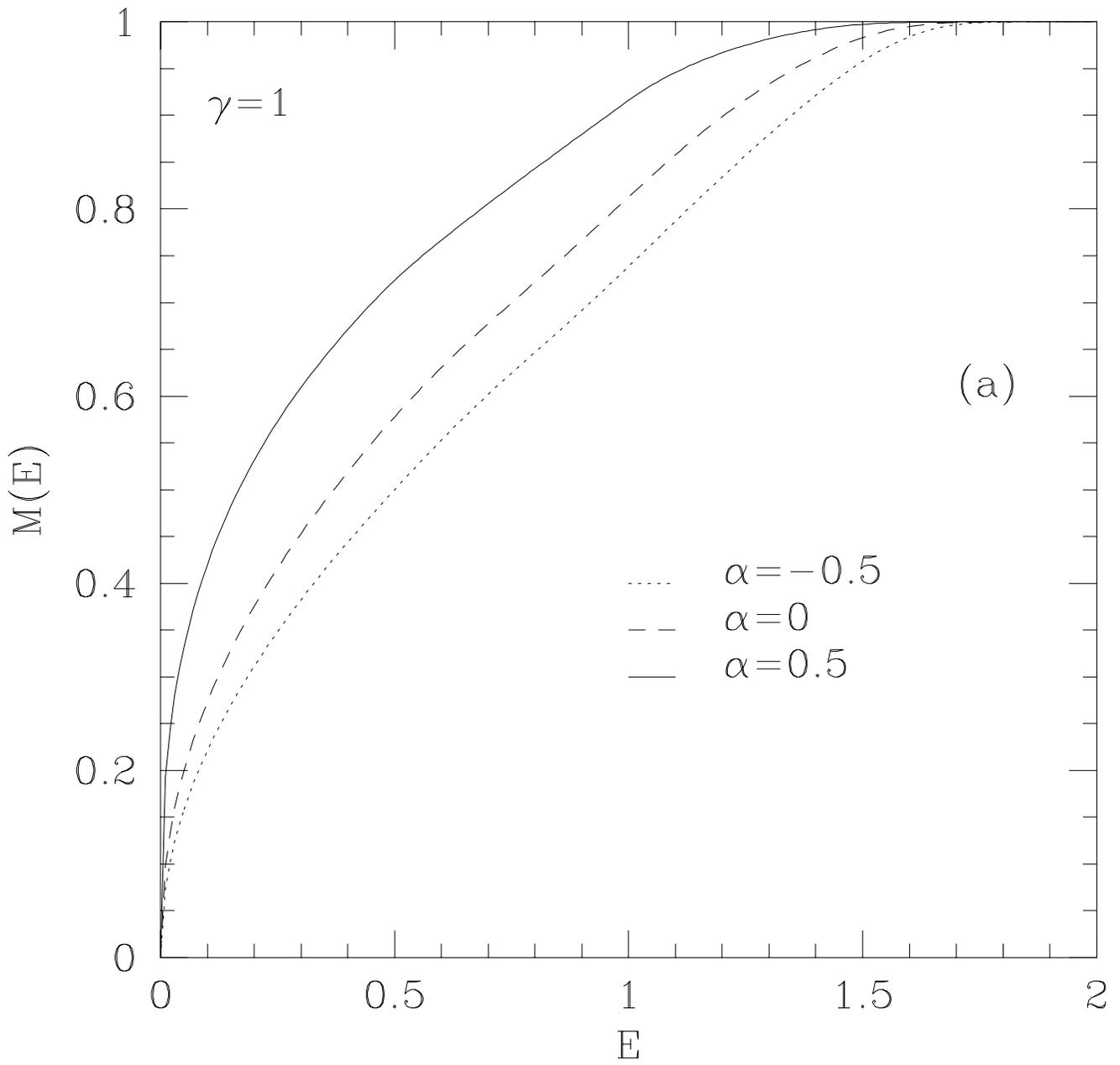
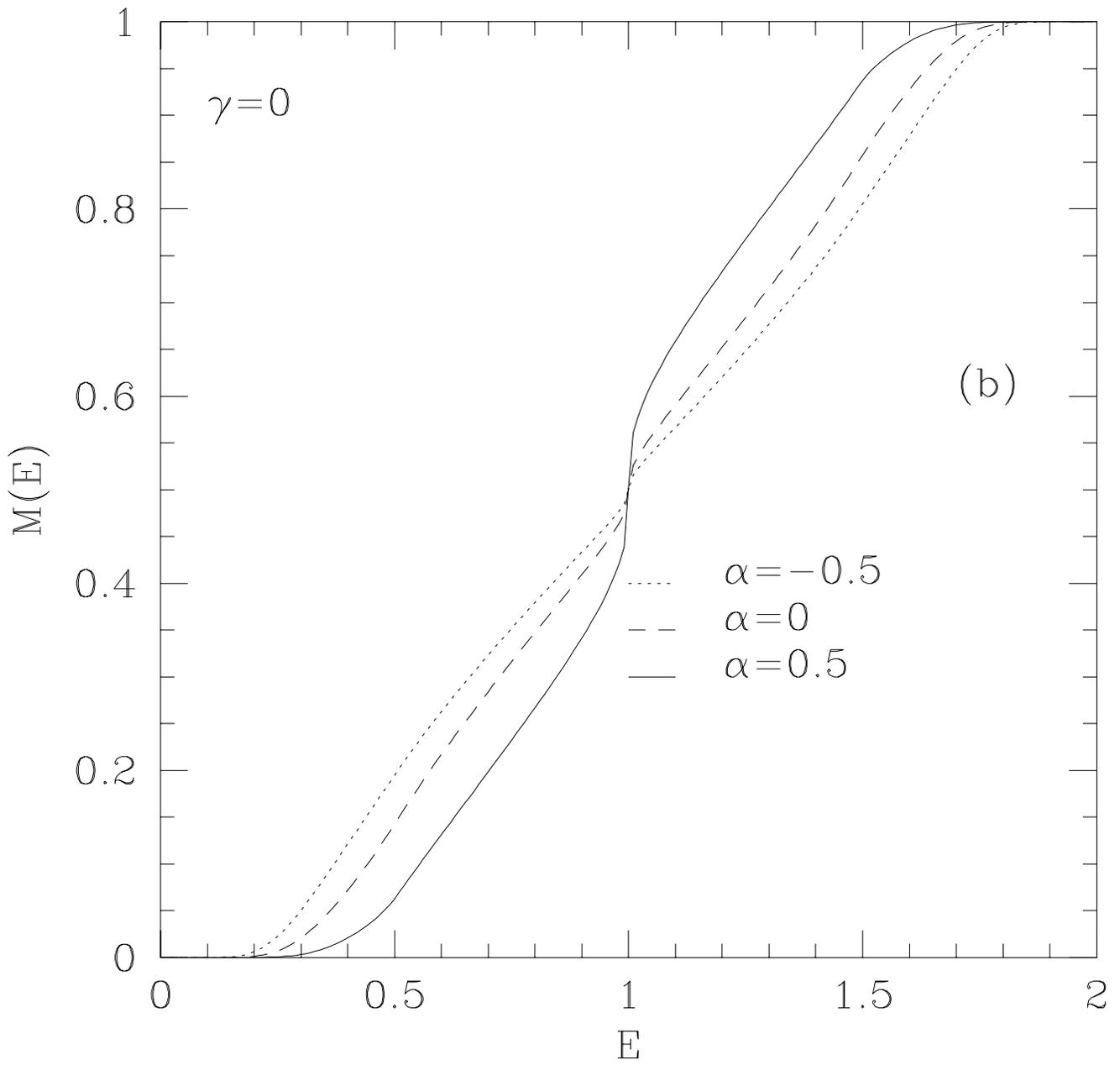
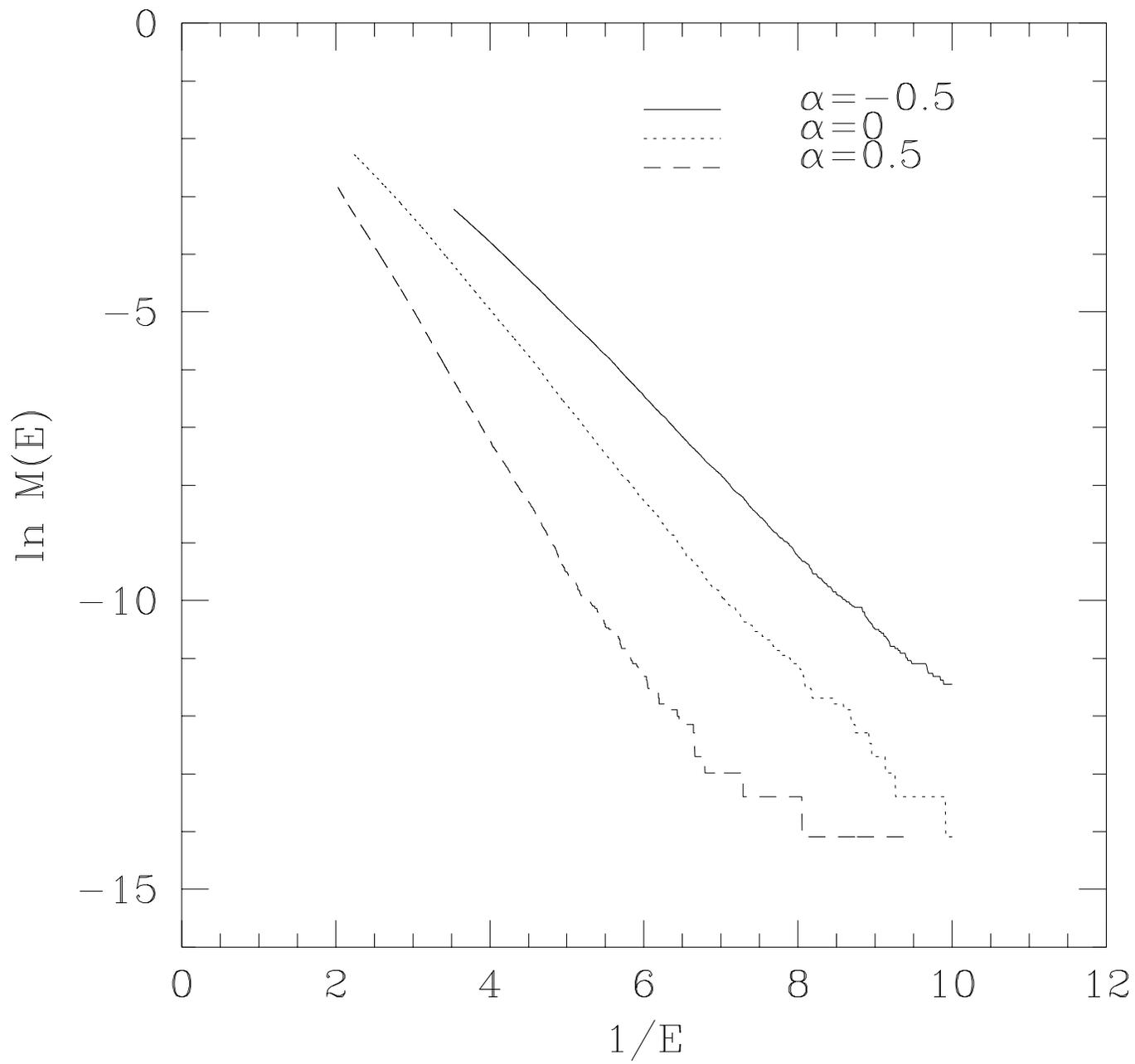
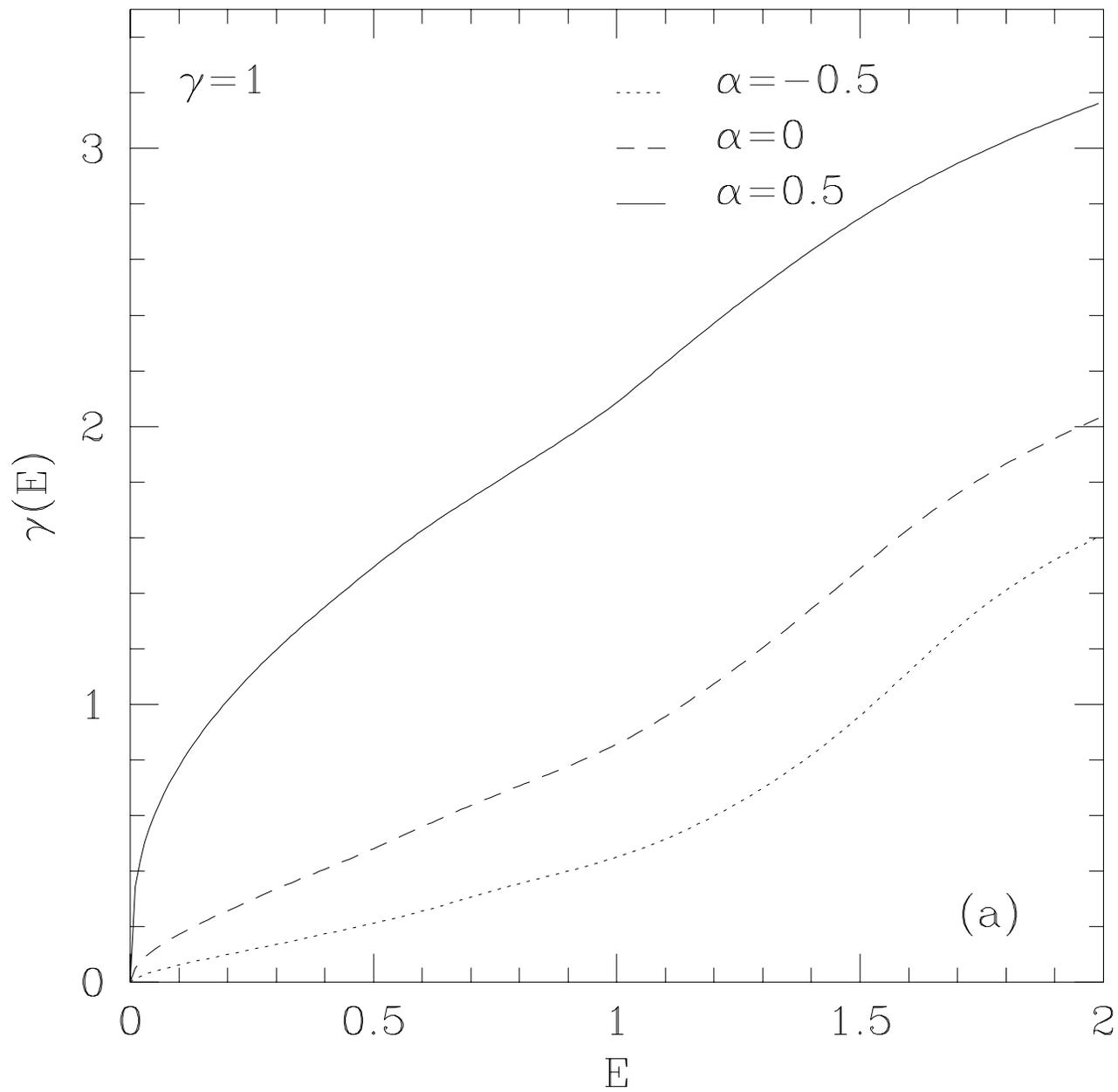
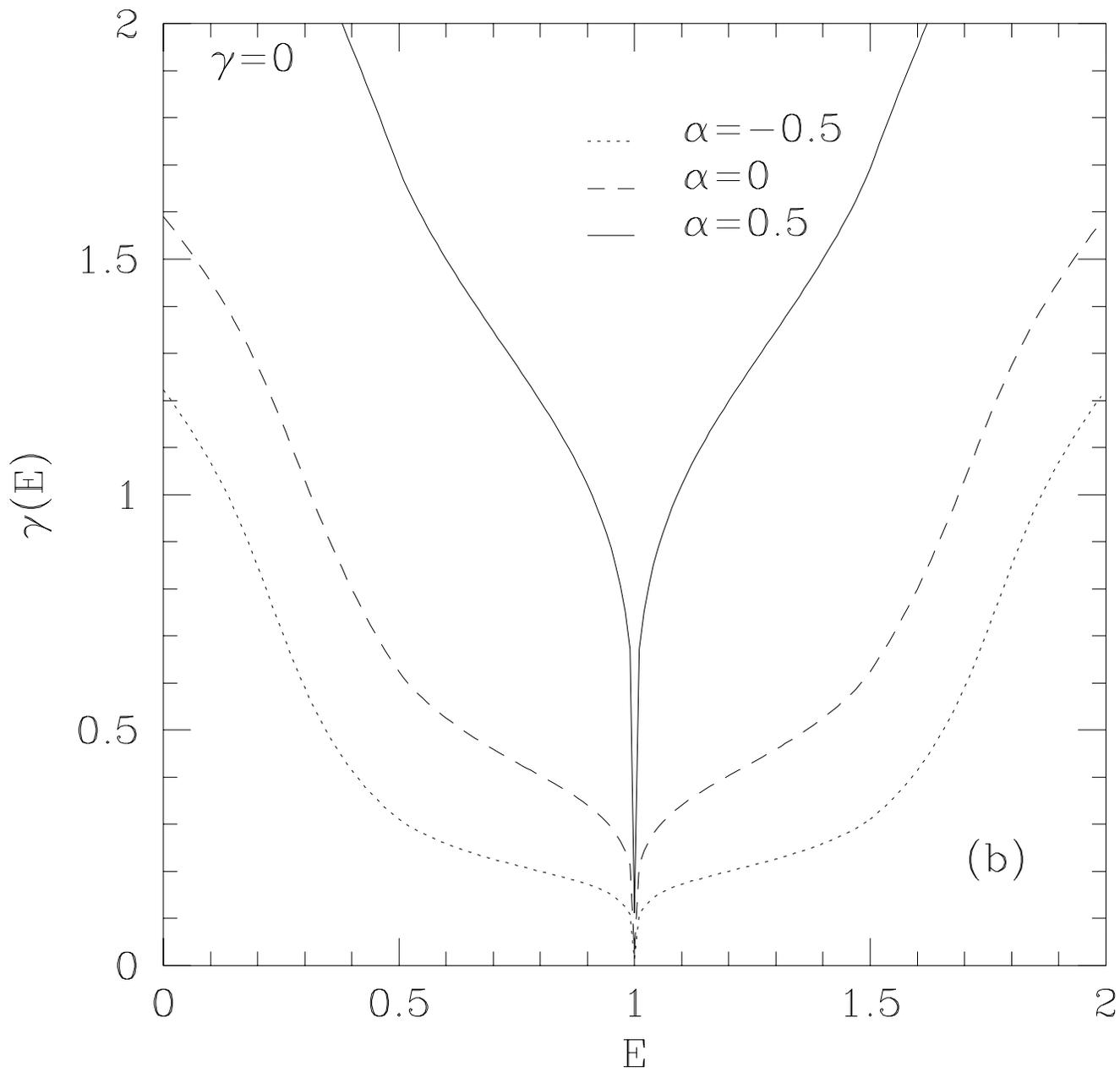
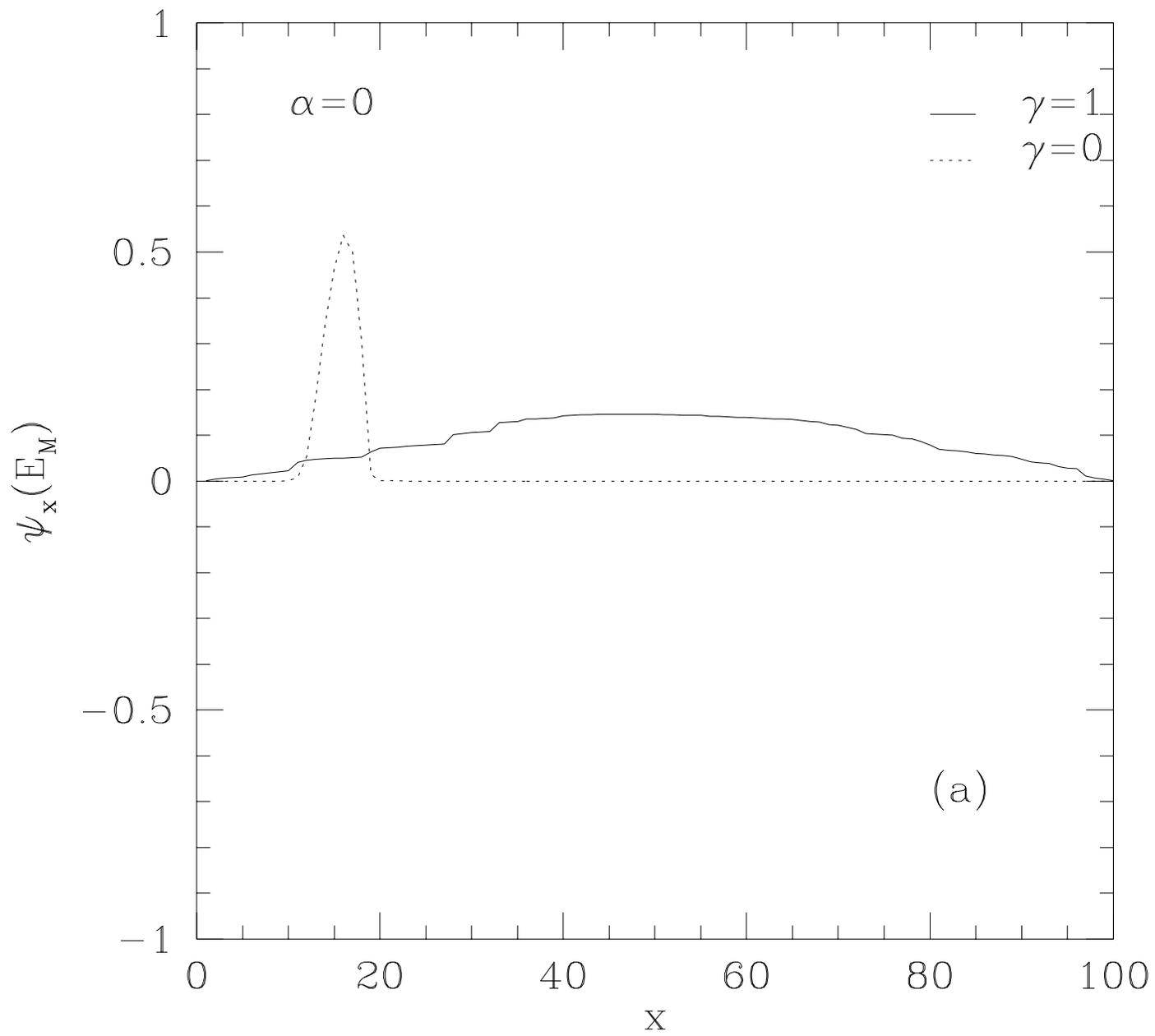
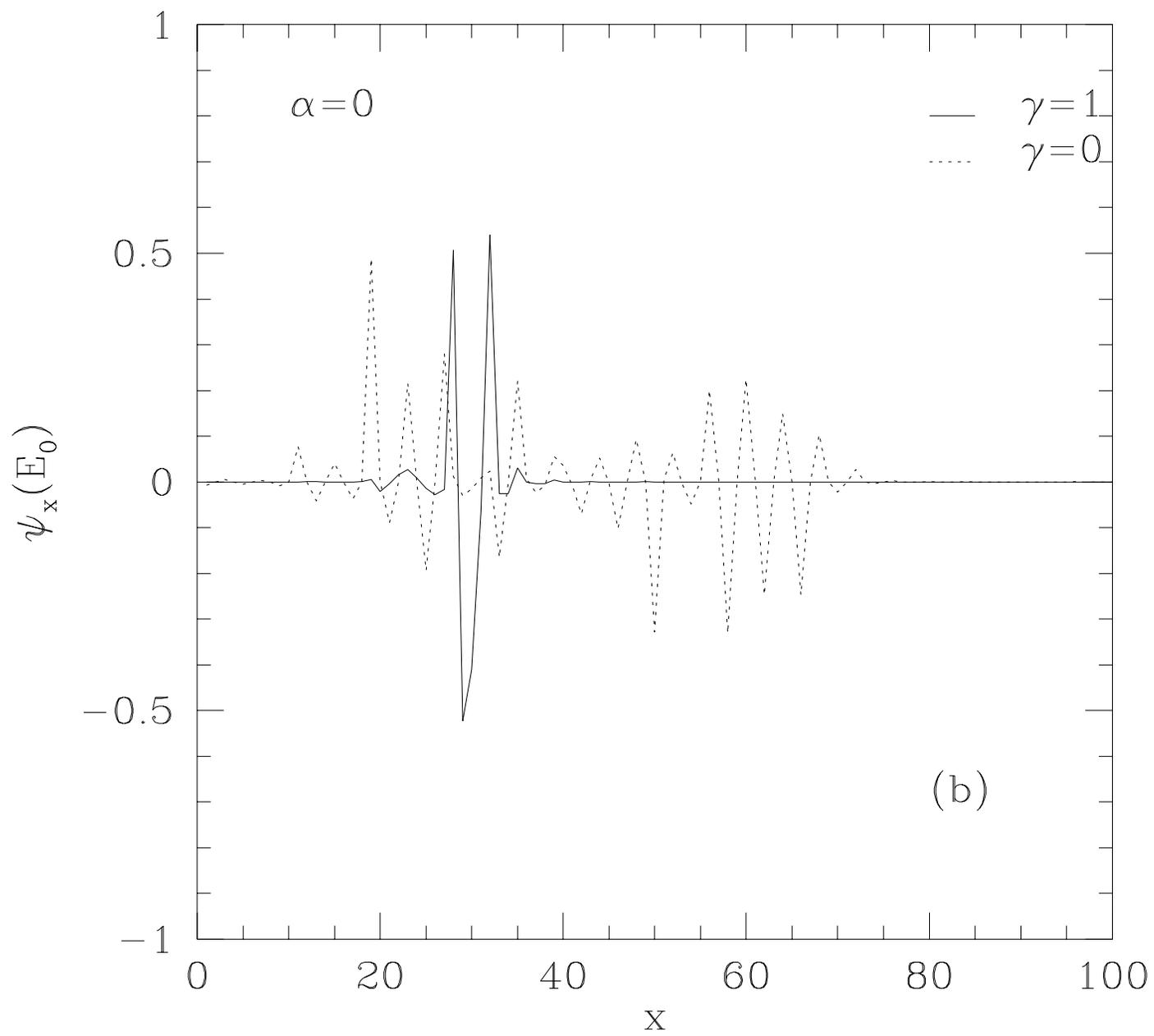
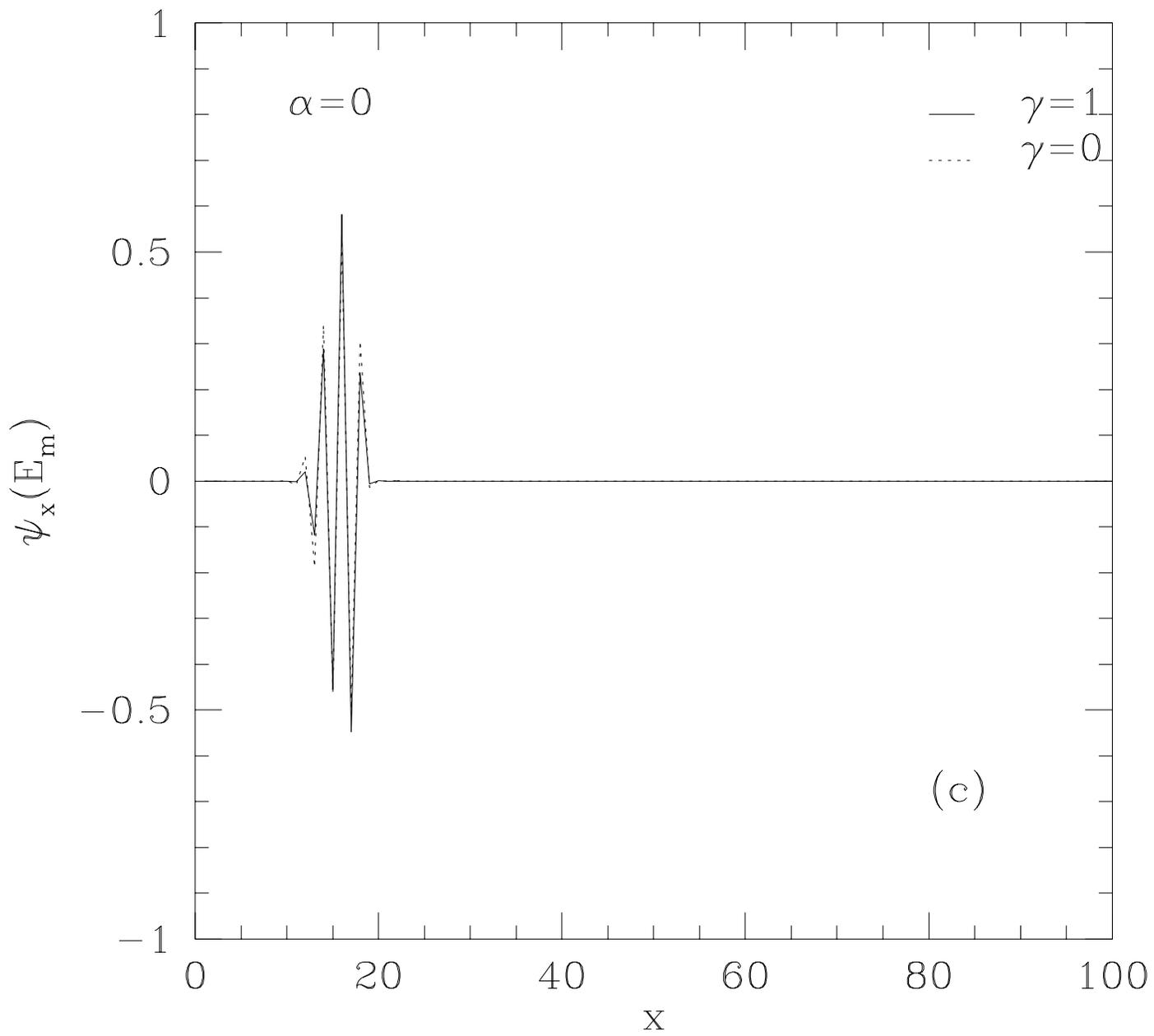
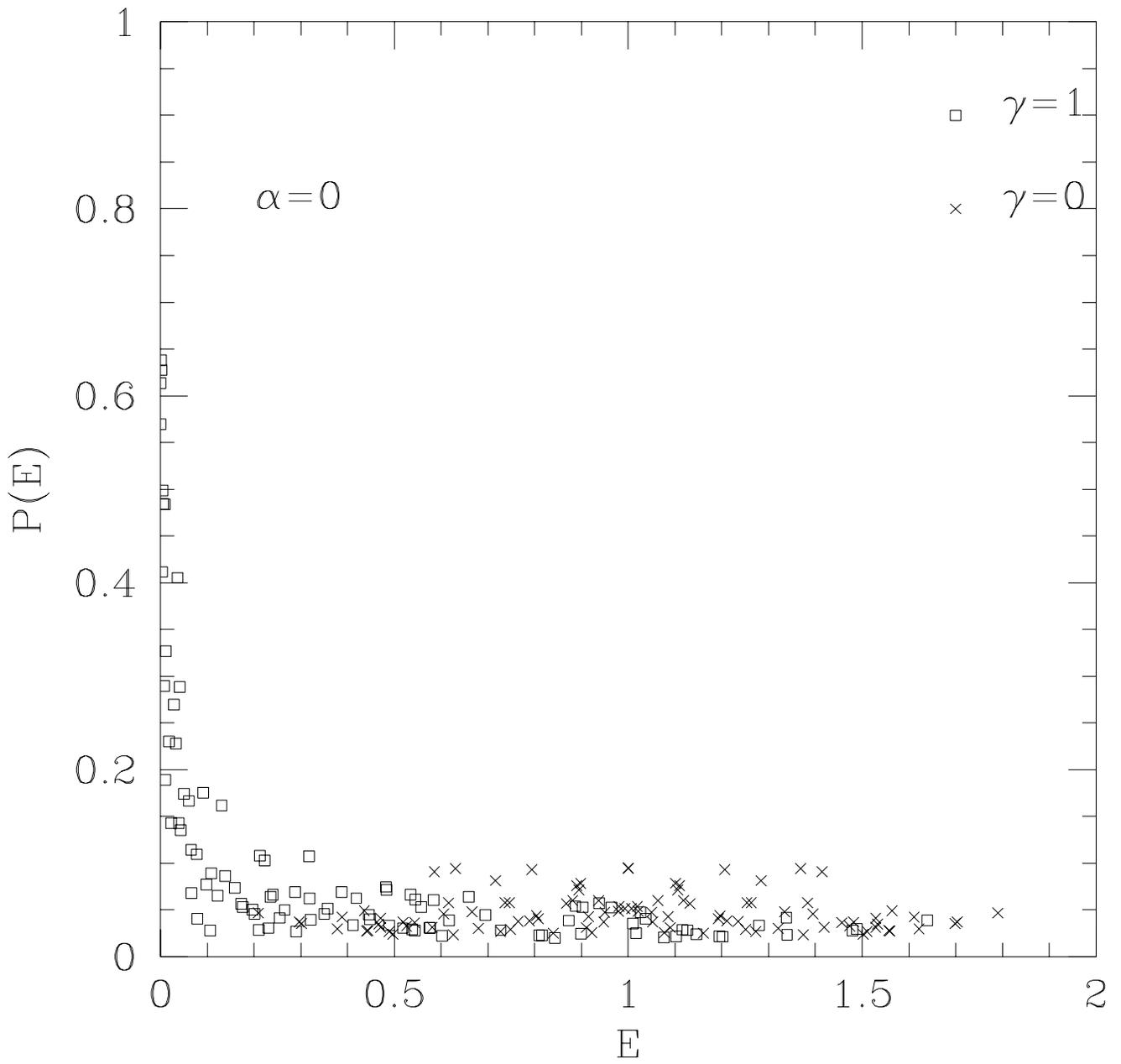
\end{document}